\documentclass{article}

\usepackage{mlps_conf}
\usepackage{times}
\usepackage{epsfig}
\usepackage{graphicx}
\usepackage{amsmath}
\usepackage{amssymb}
\usepackage{pgfplots}
\pgfplotsset{compat=1.18}
\usepackage{booktabs}
\usepackage{enumitem}

\usepackage[utf8]{inputenc} 
\usepackage[T1]{fontenc}    
\usepackage{url}            
\usepackage{booktabs}       
\usepackage{amsfonts}       
\usepackage{nicefrac}       
\usepackage{microtype}      
\usepackage{fontawesome5}
\usepackage{booktabs}
\usepackage{tabularx}
\usepackage{subcaption}
\usepackage{pifont} 
\definecolor{myGray}{rgb}{0.5, 0.5, 0.5}
\definecolor{myRed}{rgb}{0.808,0.067,0.149}
\definecolor{myGreen}{rgb}{0.067,0.708,0.149}
\usepackage{multirow}

\usepackage{longtable}
\usepackage[skins]{tcolorbox} 
\tcbuselibrary{breakable}
\usepackage{float}
\newfloat{Prompt}{htbp}{loa}
\usepackage{placeins}
\usepackage{times}
\usepackage{epsfig}
\usepackage{graphicx}
\usepackage{amsmath}
\usepackage{amssymb}
\definecolor{darkgreen}{rgb}{0.0,0.5,0.0}
\usepackage{booktabs, xcolor, colortbl}
\usepackage{enumitem}

\usepackage[rightcaption]{sidecap}
\usepackage{graphicx}

\copyrightnotice{U.S.\ Government work not protected by U.S.\ copyright}

\copyrightnotice{979-8-3503-2411-2/25/\$31.00 {\copyright}2025 Crown}

\copyrightnotice{979-8-3503-2411-2/25/\$31.00 {\copyright}2025 European Union}

\copyrightnotice{979-8-3503-2411-2/25/\$31.00 {\copyright}2025 IEEE}

\toappear{2025 IEEE International Workshop on Machine Learning for Signal Processing, Aug.\ 31-- Sep.\ 3, 2025, Istanbul, Turkey}

\usepackage[pagebackref=true,breaklinks=true,letterpaper=true,colorlinks,bookmarks=false]{hyperref}

\name{
David Robinson$^{1}$ \quad
Animesh Gupta$^{1}$ \quad
Rizwan Qureshi$^{1}$ \quad
Qiushi Fu$^{2}$ \quad
Mubarak Shah$^{1}$
}
\address{
$^{1}$Center for Research in Computer Vision, University of Central Florida \\
$^{2}$Mechanical and Aerospace Engineering, University of Central Florida
}

\begin{document}

\title{StrokeVision-Bench: A Multimodal Video and 2D Pose Benchmark for Tracking Stroke Recovery}

\maketitle

\begin{abstract}

Despite advancements in rehabilitation protocols, clinical assessment of upper extremity (UE) function after stroke largely remains subjective, relying heavily on therapist observation and coarse scoring systems. This subjectivity limits the sensitivity of assessments to detect subtle motor improvements, which are critical for personalized rehabilitation planning. Recent progress in computer vision offers promising avenues for enabling objective, quantitative, and scalable assessment of UE motor function. Among standardized tests, the Box and Block Test (BBT) is widely utilized for measuring gross manual dexterity and tracking stroke recovery, providing a structured setting that lends itself well to computational analysis. However, existing datasets targeting stroke rehabilitation primarily focus on daily living activities and often fail to capture clinically structured assessments such as block transfer tasks. Furthermore, many available datasets include a mixture of healthy and stroke-affected individuals, limiting their specificity and clinical utility. To address these critical gaps, we introduce \texttt{StrokeVision-Bench}, the first-ever dedicated dataset of stroke patients performing clinically structured block transfer tasks. \texttt{StrokeVision-Bench} comprises 1,000 annotated videos categorized into four clinically meaningful action classes, with each sample represented in two modalities: raw video frames and 2D skeletal keypoints. We benchmark several state-of-the-art video action recognition and skeleton-based action classification methods to establish performance baselines for this domain and facilitate future research in automated stroke rehabilitation assessment.

\end{abstract}

\begin{figure*}[t] 
    \centering
    \includegraphics[width=\textwidth]{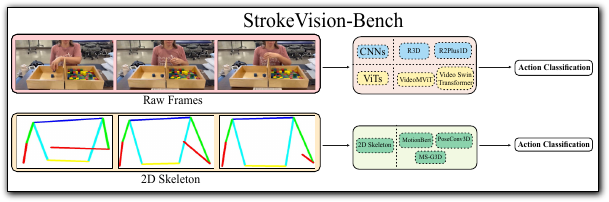} 
    \caption{Overview of \texttt{StrokeVision-Bench}. We collect 1K short videos of stroke patients performing the block‐transfer test and extract two modalities: (top) Raw RGB frames and (bottom) 2D skeletal joint trajectories. Each modality is fed into a dedicated encoder, either a video-based model (e.g., CNN or Vision Transformer) or a skeleton-based network, for action classification.}
    \label{fig:fig1_main}
\end{figure*}

\section{Introduction}
\label{sec:section1}

Stroke is the leading cause of serious chronic physical disability in the United States, impacting millions of individuals each year~\cite{update2017heart}. Among stroke survivors, approximately 95\% experience upper extremity (UE) dysfunction~\cite{gowland1992agonist}, with 30 to 66\% exhibiting a significantly impaired ability to use the affected arm~\cite{van1999forced}. Standardized routine Outcome Measures (OMs) of UE impairment are critical for informing clinical decisions about rehabilitation protocols and for tracking the progression of sensorimotor deficits~\cite{potter2011outcome,sullivan2011outcome}. However, commonly used OMs often fail to provide sufficient evidence to help therapists develop and adapt personalized care plans. This is because existing assessment tools rely on subjective grading of movement quality using a few discrete levels (e.g., Action Research Arm Test)~\cite{brunner2024external}, or the number of blocks transferred during a fixed interval~\cite{stinear2019prediction}.  As a result, these OMs lack the sensitivity needed to detect behavioural changes or to directly target specific functional deficits. 

\par To address these shortcomings, considerable research has focused on developing wearable assessment tools that measure hand and arm kinematics~\cite{o2022wearable}. However, these technologies are not widely adopted in clinical settings due to their high cost and the need for specialized training for clinical staff. To overcome these challenges in the research, we aim to develop a solution based entirely on computer vision techniques that are affordable and easily deployable in real-world clinical environments. Our goal is to develop a computer vision–based diagnostic system that runs on mobile devices, enabling objective progress tracking without costly equipment or clinical expertise~\cite{mainali2021machine}.

Video action understanding remains a fundamental and challenging problem in computer vision, requiring models to recognize and classify complex human actions from dynamic visual content~\cite{hutchinson2021video, thawakar2024mobillama}. Video action understanding requires learning rich spatio-temporal features from frame sequences, capturing human interactions, object dynamics, and overall context \cite{qureshi2025thinking, raza2025responsible}. Transformer-based vision encoders (ViTs)\cite{dosovitskiy2020image, narnaware2025sb} have emerged as the gold standard for modeling spatio-temporal information and capturing global contextual relationships in visual data. Building on this foundation, Swin Transformers\cite{liu2021swin} enhance the standard self-attention mechanism by introducing a hierarchical shifted window approach, enabling more efficient and scalable learning of spatial representations. In the clinical setting, using the 2D skeleton serves two purposes: (1) it provides rich information about the mobility of different parts of the body. (2) it does not reveal the privacy attributes of the patients \cite{vayani2024all}. We expanded our benchmarking setup upon advances in skeleton-based representations and evaluated MotionBert~\cite{motionbert2022}, PoseConv3D~\cite{duan2022revisiting}, and MS-G3D~\cite{liu2020disentangling} on our proposed \texttt{StrokeVision-Bench}.

Despite the strong performance of 2D skeleton action classification methods, their effectiveness and efficiency on medical datasets remain open questions. Building on this motivation, we introduce \texttt{StrokeVision-Bench}, a clinical video dataset consisting of human-centric recordings. These videos capture patients performing the task of transferring blocks from one location to another, recorded both before and after the clinical sessions. The individuals featured in the dataset suffer from upper extremity (UE) dysfunction and regularly attend rehabilitation sessions for weekly physical treatment aimed at improving the mobility of their body parts. To support this objective, it is necessary to monitor the individual's progress across multiple weeks. Two types of information are particularly valuable for doctors: (1) the total number of objects transferred by the individual before and after the sessions, and (2) the joint angle between the shoulder and abdomen, which provides insight into improvements in the body movement. We benchmark existing state-of-the-art video action and 2D skeletons classification methods in this study. Our dataset is annotated, where each video is labelled into one of four categories: (i) Grasping, (ii) Non-task Movement, (iii) Transport with Object, (iv) and Transport without Object. 

To summarize, we make the following key contributions in this work:
\begin{itemize}

    \item We introduced \texttt{StrokeVision-Bench}, the first-ever dataset of 1000 videos covering four subactions of the clinical box and block test for computer vision based analysis. Each video includes both raw RGB frames and corresponding 2D skeleton poses, computed using Sapiens~\cite{khirodkar2024sapiens} to enable fine‐grained analysis of patient movement. By including recordings before and after the sessions, our dataset captures changes in movement speed across different body parts, which is essential to assess rehabilitation progress, and also, training models to detect subtle improvements in mobility.

    \item \texttt{StrokeVision-Bench} is the first stroke rehabilitation benchmarking dataset that focuses exclusively on stroke patients, filling a gap in existing work and enabling the systematic evaluation of different methods for tracking patient progress.

    \item We evaluated four video action classification methods, namely R3D\cite{tran2018closer}, R2Plus1D\cite{tran2018closer}, Video MViT~\cite{li2021mvitv2}, and Video Swin Transformer~\cite{liu2022video}. Additionally, we evaluated 2D skeleton-based action classification methods, including MotionBERT~\cite{motionbert2022}, PoseConv3D~\cite{duan2022revisiting}, and MS-G3D~\cite{liu2020disentangling}. We expect that this dataset will further advance research in automated stroke-prediction. 

\end{itemize}



\begin{table*}[t]
\centering
\setlength{\tabcolsep}{8pt}
\renewcommand{\arraystretch}{1.2}
\resizebox{0.85\textwidth}{!}{%
\begin{tabular}{lccc}
    \toprule
    \textbf{Model} & \textbf{Pretrained Weights} & \textbf{Modality} & \textbf{Accuracy (\%)} \\
    \midrule
    R3D~\cite{tran2018closer}                   & \textit{Kinetics-400} & Frames      & 87.68 $\pm$ 2.48 \\
    R2Plus1D~\cite{tran2018closer}              & \textit{Kinetics-400} & Frames      & 86.96 $\pm$ 1.20 \\
    Video MViT~\cite{li2021mvitv2}              & \textit{Kinetics-400} & Frames      & 77.14 $\pm$ 2.15 \\
    Video Swin Transformer~\cite{liu2022video}  & \textit{Kinetics-400} & Frames      & 78.93 $\pm$ 2.15 \\
    MotionBERT~\cite{motionbert2022}            & \textit{NTU RGB+D}    & 2D Joints   & 84.29 $\pm$ 1.35 \\
    PoseConv3D~\cite{duan2022revisiting}        & \textit{NTU RGB+D}    & 2D Joints   & 68.93 $\pm$ 3.06 \\
    MS-G3D~\cite{liu2020disentangling}          & None                  & 2D Joints   & 78.39 $\pm$ 1.72 \\
    \bottomrule
\end{tabular}%
}
\vspace{-0.5em}
\caption{Accuracy (mean $\pm$ standard deviation) of different models on \texttt{StrokeVision-Bench}. Frame-based methods (R3D, R2Plus1D, Video MViT, Video Swin Transformer) are pretrained on \textit{Kinetics-400}, while skeleton-based methods (MotionBERT, PoseConv3D, MS-G3D) use \textit{NTU RGB+D} or no pretraining.}
\label{tab:methods_comparison}
\vspace{-1.5em}
\end{table*}

\section{Related Work}
Video-based action recognition has shown remarkable success in modeling complex human activities in everyday environments~\cite{lo2024predictive, mahmood2024vurf, raza2025responsible}. Translating these techniques to medical domains, particularly stroke rehabilitation, demands curated datasets and models capable of handling clinically meaningful movements and patient variability. \\

\textbf{Expensive and training-intensive methods:} Existing approaches are costly and require specialized clinician training. Previous methods~\cite{wade2010automated, wade2011virtual, hebert2012case, gupta2023data, balasubramanian2009robot, chen2013discriminant} have developed technologies to measure hand and arm kinematics in individuals with sensorimotor impairments. However, these methods rely on multiple wearable sensors or expensive motion capture systems, making them costly and time-consuming to deploy. They also require customized equipment designed for laboratory-based tasks that are unfamiliar to clinicians~\cite{balasubramanian2009robot, gupta2025play, dukelow2010quantitative, vayani2025all}, further increasing the need for specialized training. In many cases, the complexity of setup and interpretation limits their adoption outside of controlled research environments. Moreover, such systems often fail to generalize well to in-home or community-based settings where real-world rehabilitation occurs. To overcome these challenges, we propose a new approach based on video action classification that uses low-cost cameras to track and record patient mobility, removing the dependency on wearable devices.


\textbf{Existing video action classification datasets for stroke:} Existing video action classification datasets for stroke rehabilitation are limited. StrokeRehab \cite{kaku2022strokerehab} offers large‐scale recordings of daily activities, such as brushing and combing, collected from 20 healthy individuals and 31 stroke‐impaired patients. Another study \cite{parnandi2023data} uses videos of healthy subjects to identify and quantify movement abnormalities, but models trained exclusively on healthy data suffer substantial performance degradation when applied to stroke‐impaired videos. By contrast, our \texttt{StrokeVision-Bench} centers on a standardized clinical task: we collect video of each patient performing the Box and Block Test both before and after rehabilitation sessions, thereby enabling direct evaluation of changes in motor function.

\textbf{Existing video and 2D skeleton action classification methods:} Convolutional Neural Networks (CNNs) and Vision Transformers (ViTs) have both been widely applied to video action recognition \cite{saeed2025beyond}. Duan et al.~\cite{duan2022revisiting} proposed spatio‐temporal CNN architectures, and Li et al.~\cite{li2021mvitv2} together with Liu et al.~\cite{liu2021swin} introduced transformer‐based models that capture long‐range dependencies across frames. For skeleton‐based classification, MotionBERT~\cite{motionbert2022} uses 2D joint sequences but relies on accurate pose estimates that may degrade in clinical recordings, while PoseConv3D~\cite{duan2022revisiting} attains high accuracy on controlled benchmarks but has not been evaluated on patient videos. MS‐G3D~\cite{liu2020disentangling} achieves strong results on standard action datasets but depends on dense annotations that are costly to obtain in medical settings. 


\section{StrokeVision Benchmark}


\begin{figure*}[t]
  \centering
  \begin{subfigure}[b]{0.75\textwidth}
    \includegraphics[width=\textwidth]{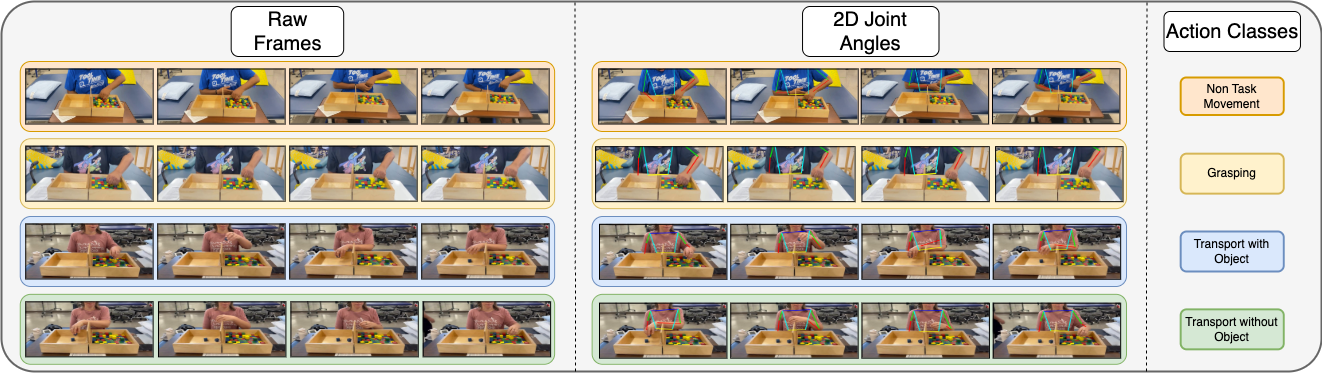}
    \caption{Visualization of raw RGB frames alongside overlaid 2D skeleton keypoints, annotated with their corresponding action classes.}
    \label{fig:image_curation1}
  \end{subfigure}
  \hfill
  \begin{subfigure}[b]{0.24\textwidth}
  \centering
  \makebox[\textwidth][c]{%
    \resizebox{0.88\textwidth}{!}{




\begin{tikzpicture}
  \begin{axis}[
      ybar=0pt,
      bar width=18pt,
      symbolic x coords={Non task\\movement, Grasping, Transport\\with object, Transport\\without object},
      xtick=data,
      ymin=0, ymax=400,
      ytick={0,50,...,400},
      tick label style={font=\small},
      xticklabel style={align=center,font=\small},
      ymajorgrids=true,
      grid style={gray!30},
      nodes near coords,
      every node near coord/.append style={font=\small},
      legend style={
        at={(rel axis cs:1,1)},
        anchor=north east,
        font=\small,
        draw=black,
        fill=white,
        /tikz/every even column/.append style={column sep=0.5cm}
      },
      cycle list={{fill=blue!50},{fill=red!50}}
  ]
    \addplot+[] coordinates {
      (Non task\\movement,148) (Grasping,372)
      (Transport\\with object,100) (Transport\\without object,159)
    };
    \addplot+[] coordinates {
      (Non task\\movement,38) (Grasping,93)
      (Transport\\with object,26) (Transport\\without object,40)
    };
    \legend{Train, Validation}
  \end{axis}
\end{tikzpicture}}%
  }
    \caption{Train and validation sample counts for each action category in \texttt{StrokeVision-Bench}.}
    \label{fig:image_curation2}
  \end{subfigure}
  \vspace{-0.5em}
  \caption{Overview of StrokeVision-Bench: (a) shows example raw‐frame and 2D‐skeleton modalities with action labels, while (b) presents the distribution of training and validation samples across the four action classes.}
  \label{fig:side_by_side}
  \vspace{-1.5em}
\end{figure*}
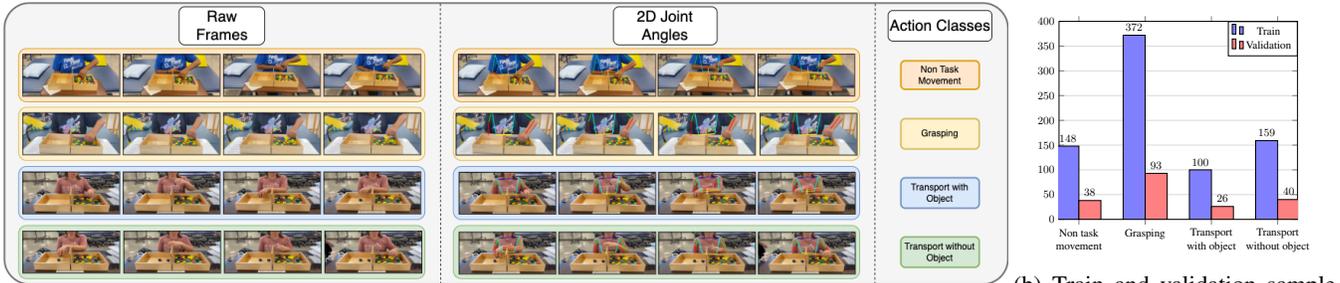

\subsection{Curation of StrokeVision}

We collected videos of patients, where each patient is performing the task of moving a block from one box to another. We curated videos recorded both pre- and post-session, enabling evaluation of the network’s ability to understand motion differences. Ideally, a patient demonstrates improved mobility after the session, resulting in a noticeable difference in motion speed between the two recordings. Each video was manually annotated into \textit{four} action classes. \texttt{StrokeVision} includes a total of 1000 videos, each with a duration of 1 second and containing 30 frames. Curating an equal number of videos of distinct actions is challenging because stroke-impaired patients often struggle to transfer the block between boxes, resulting predominantly in grasping or non-task movements (Fig.~\ref{fig:image_curation2}). Additionally, we created train-test splits while ensuring there was no data leakage between them. For more details regarding the number of videos in each action class, refer to Fig.~\ref{fig:image_curation2}, and for visualizations of our videos in \texttt{StrokeVision}, refer to Fig.~\ref{fig:image_curation1}.

\subsection{Benchmarking Setup}
As detailed in Section~\ref{sec:section1}, our goal is to compare state-of-the-art video action and 2D skeleton-based recognition methods. We denote an input video as: 
\vspace{-1em}
\[
V \in \mathbb{R}^{T \times C \times H \times W},
\] 
where \(T\) is the frame count, \(C\) the number of channels, and \(H\) and \(W\) the frame height and width.  Each video is classified into one of four action categories using either raw frames or 2D skeleton keypoints. To obtain 2D skeleton data, we apply Sapiens~\cite{khirodkar2024sapiens} a state-of-the-art model by Meta, yielding 
\[
V_{J}\in \mathbb{R}^{T \times J \times 2},
\] 
where \(J=17\) keypoints and the last dimension represents the \((x,y)\) coordinates.  


\section{Experiments}


\textbf{Models:} We examined multiple CNN and transformer-based networks for action classification using raw frames. For CNNs, we evaluated residual-based network architectures introduced in \cite{tran2018closer}, including 3D ResNet (referred to as R3D), and ResNet with (2+1)D convolutions (referred to as R(2+1)D). For vision transformers, we selected Video MViT~\cite{li2021mvitv2}, and Video Swin Transformer~\cite{liu2022video}. For 2D keypoint-based methods, we evaluated MotionBert~\cite{motionbert2022}, PoseConv3D~\cite{duan2022revisiting}, and MS-G3D~\cite{liu2020disentangling}. We initialized all frame-based networks with weights pretrained on the Kinetics-400 dataset. For 2D skeleton-based models, we used weights pretrained on NTU RGB+D, with the exception of MS-G3D because its pretraining employed a different set of skeleton keypoints.



\textbf{Experimental Setup:} We evaluated several models finetuned on our proposed \texttt{StrokeVision-Bench} using \textit{accuracy} as the evaluation metric. All our experiments are implemented in PyTorch~\cite{paszke2017automatic} and executed on a single NVIDIA V100 GPU with 32 GB of VRAM. 

\section{Benchmarking on StrokeVision-Bench}
\begin{figure*}[t]
  \centering
  \includegraphics[width=0.9\linewidth]{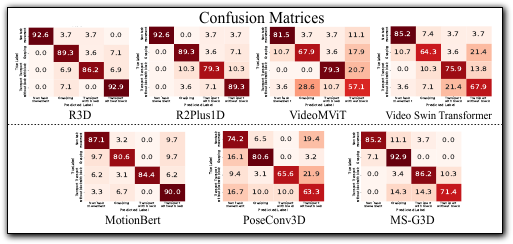}
  \caption{Confusion matrices for seven models (R3D, R2Plus1D, Video MViT, Video Swin, MotionBert, PoseConv3D, MS-G3D) on four \texttt{StrokeVision-Bench} actions. Most models excel in “Non-task movement” and “Grasping” but show varying confusion between “Transport with block” and “Transport without block.”}
  \label{fig:confusion_matrices}
  \vspace{-1em}
\end{figure*}


\textbf{Overall Results:} We evaluated four video-based models on raw frames and three skeleton-based methods on \texttt{StrokeVision-Bench}, reporting the results in Table \ref{tab:methods_comparison}. First, convolutional networks outperform vision transformers on our small and challenging dataset. For instance, accuracy falls from 87 \% for R3D~\cite{tran2018closer} to 74\% for the Video Swin Transformer~\cite{liu2021swin}, reflecting the transformers’ greater data requirements for learning effective global representations. Second, among skeleton-based approaches, MotionBert~\cite{motionbert2022} achieves the highest accuracy at 84\%.

\textbf{Raw frames or 2D Skeleton?} We analyze the performance of two modalities, raw video frames and 2D skeletons, in our \texttt{StrokeVision-Bench}. Overall, the R3D~\cite{tran2018closer} model achieves the highest accuracy at 87\%, outperforming MotionBert’s~\cite{motionbert2022} 84\%. For clinical applications focused solely on quantifying block transfers before and after treatment, frame-based CNNs such as R3D are the best choice. If the clinic also wishes to assess improvements in patients’ overall mobility, skeleton-based methods such as MotionBert are preferable, since it deliver nearly equivalent accuracy while providing joint-level movement data that can be used to evaluate mobility improvements.

\textbf{Vision Transformers vs.\ 2D Skeleton Methods:} Our experiments confirm that vision transformers require large datasets to achieve strong performance, as demonstrated in DeiT~\cite{pmlr-v139-touvron21a}. In clinical settings, where data curation is challenging, transformers are therefore a suboptimal choice. In contrast, 2D skeleton–based methods learn robust representations even from limited data and consistently outperform vision transformers. Furthermore, skeleton–based approaches preserve patient privacy attributes and provide detailed, joint‐level mobility information.

\textbf{Confusion Matrices Analysis:} Across all seven models, “Non-task movement” is easy to recognize, with most architectures exceeding 85\% accuracy, while “Transport with block” versus “Transport without block” remains challenging (Fig. \ref{fig:confusion_matrices}). The 3D CNNs R3D and R2Plus1D show uniform performance, approximately 92\% on Non-task movement, 89\% on Grasping, and 86–93\% on transport actions, while attention-based models Video MViT and Video Swin Transformer suffer in classifying the two transport variants (only 57.1\% and 67.9\% correct on “without block”). The 2D skeleton-based networks takes a middle ground: MotionBert achieves CNN-level accuracy (87.1\% Non-task, 80.6\% Grasping, 84.4\% and 90.0\% transports), PoseConv3D boosts Grasping (80.6\%) but struggles on Non-task (74.2\%) and both transports (63–66\%), and MS-G3D offers the most balanced pose-driven performance (92.9\% Grasping, 86.2\% with-block, 71.4\% without-block) while still finding “transport without block” the complex case.


\section{Limitations}

While \texttt{StrokeVision-Bench} represents a significant advance in benchmarking action recognition methods for stroke patients, it has several limitations. First, the dataset includes only the Box and Block Test, leaving other clinical assessments of patient mobility unexamined. Second, the current dataset size is limited and would benefit from expansion through partnerships with clinical centers.
\section{Conclusion}

In this paper, we introduce \texttt{StrokeVision-Bench}, a new dataset of 1000 videos documenting stroke rehabilitation patients performing the Box and Block Test. We introduce the first dataset exclusively for stroke-impaired patients performing the standard test. By carefully curating recordings made before and after rehabilitation sessions, we capture the changes in motion speed that indicate functional improvement. The dataset includes both raw-frame and 2D-skeleton modalities, allowing evaluation of convolutional neural networks, vision transformers and skeleton-based methods. Despite the dataset's small size and inherent challenges, CNN-based models achieve the highest overall accuracy. Skeleton-based approaches perform comparably while preserving patient privacy and providing detailed insights into joint-level mobility.



{\small
\bibliographystyle{ieee_fullname}
\bibliography{egbib}
}

\end{document}